\documentclass{amsart}
\usepackage{amsmath, amsthm, amscd, amsfonts, amssymb, graphicx, color}
\usepackage{graphicx}
\usepackage{float}
\usepackage{pdfpages}
\usepackage[bookmarksnumbered, colorlinks, plainpages]{hyperref}
\numberwithin{equation}{section}

\begin{document}

\title{ Generalized additive models to capture the death rates in Canada COVID-19}

\author{Farzali Izadi}
\address{Mathematics Department Urmia university of,
Urmia, Iran}
\email{f.izadi@utoronto.ca}



\date{June, 25, 2020.}



\begin{abstract}
To capture the death rates and strong weekly, biweekly and probably monthly patterns in the Canada COVID-19, we utilize the generalized additive models in the absence of direct statistically based measurement of infection rates. By examining the death rates of Canada in general and Quebec, Ontario and Alberta in particular, one can easily figured out that there are substantial overdispersion relative to the Poisson so that the negative binomial distribution is an appropriate choice for the analysis. Generalized additive models (GAMs) are one of the main modeling tools for data analysis. GAMs can efficiently combine different types of fixed, random and smooth terms in the linear predictor of a regression model to account for different types of effects. GAMs are a semi-parametric extension of the generalized linear models (GLMs), used often for the case when there is no a priori reason for choosing a particular response function such as linear, quadratic, etc. and need the data to 'speak for themselves'. GAMs do this via the smoothing functions and take each predictor variable in the model and separate it into sections delimited by 'knots', and then fit polynomial functions to each section separately, with the constraint that there are no links at the knots - second derivatives of the separate functions are equal at the knots.
\end{abstract}

\maketitle



\section{Introduction}

\noindent \label{intro} In late December 2019, some severe pneumonia cases of unknown cause were reported in Wuhan, Hubei province, China. These cases epidemiologically linked to a seafood wholesale market in Wuhan, although many of the first 41 cases were later reported to have no known exposure to the market. In early January 2020, the World Health Organisation (WHO) named this novel coronavirus as Severe Acute Respiratory Syndrome Coronavirus 2 (SARS-CoV-2) and the associated disease as COVID-19 \cite{W}. Following this report, there has been a rapid increase in the number of cases as on 24th June 2020, there were over 9.2 million confirmed cases and almost 475,000 deaths worldwide, while the confirmed cases in Canada was 103,000, with 8,500 deaths mostly distributed in three provinces of Quebec, Ontario, and Alberta. The death rates for these cases which are available from
\url{https://www.canada.ca/
en/public-health/services/diseases/
2019-novel-coronavirus-infection.html}
	
\noindent are shown in figure 1.
\begin{figure}
	\begin{center}
		\includegraphics[scale=0.8]{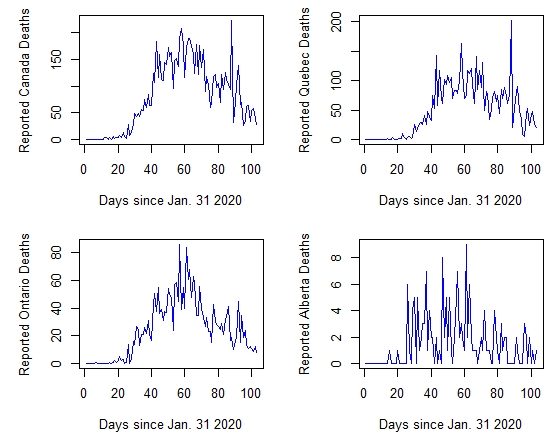}
		\caption{Reported deaths with COVID-19 in the Canada, Quebec, Ontario and Alberta since January 31, 2020
			The Canada lock down started on March 17, 2020.}\label{timeseries.jpeg}
	\end{center}
\end{figure}

\noindent From the box plot one can see that there are a few outliers in the Alberta data only, see figure 2.
\begin{figure}
	\begin{center}
		\includegraphics[scale=.7]{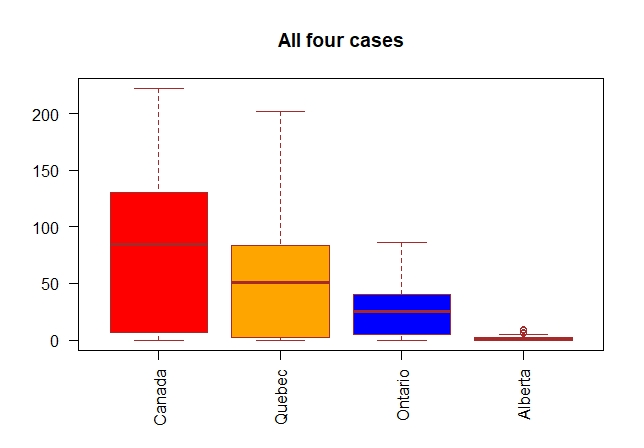}
		\caption{Note that there are some outliers in Alberta deaths.}\label{box.jpeg}
	\end{center}
\end{figure}
Also the trend, seasonal and random components of each case can be obtained by decomposition function of time series where
in figure 3 we plotted only the Canada deaths.
\begin{figure}
	\begin{center}
		\includegraphics[scale=0.7]{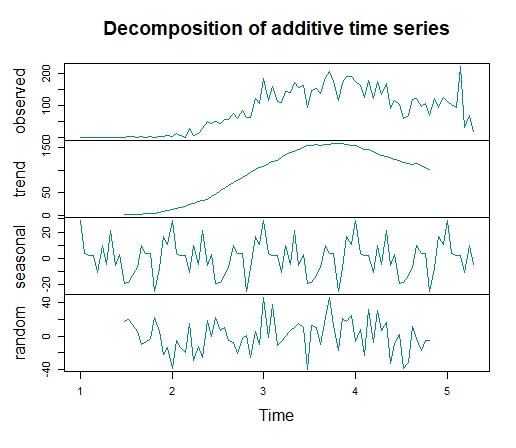}
		\caption{Decomposition of Canada death series into trend, seasonal and random components.}\label{decomp.jpeg}
	\end{center}
\end{figure}
To see the normality of the data, we can plot the histogram as well as the density of the deaths, figure 4.
The plots clearly show that there are substantial overdispersion relative to the Poisson for all cases except possibly
for the Alberta data.

\begin{figure}
	\begin{center}
		\includegraphics[scale=.8]{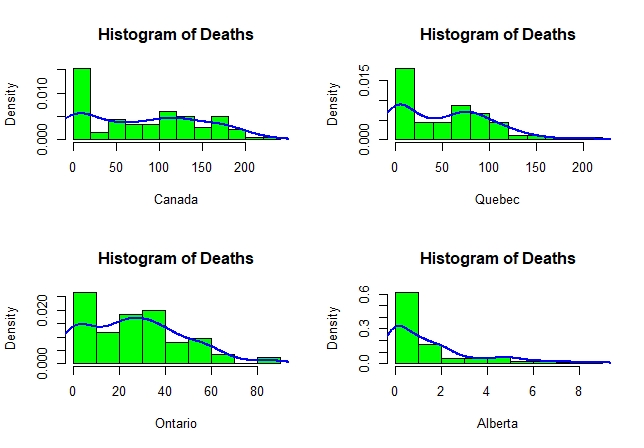}
		\caption{ Histogram and the density of the deaths series.
			Note that the substantial overdispersion in all cases.}\label{hist.jpeg}
	\end{center}
\end{figure}

\section{Generalized additive models}
\noindent In general, generalized additive models (GAMs), see e.g. S. Wood, 2017, \cite{SWB} are a semi-parametric extension of the generalized linear models (GLMs), used often for the case when there is no a priori reason for choosing a particular response function (such as linear, quadratic, etc.) and need the data to speak for themselves. GAMs first introduced by Hastie and Tibshirani, 1986 \cite{HT1} and Hastie and Tibshirani, 1990 \cite{HT2} and
are widely used in practice (Ruppert et al. 2003) \cite{REA}; (Fahrmeir, L., Lang, S. 2001) \cite{FL};
(Fahrmeir, et al. 2004) \cite{FKL1}  (Fahrmeir et al. 2013) \cite{FKL2}.
On the other hand, GLMs themselves are extension of the linear models (LMs). To understand the thing better, let us start with LMs. For a univariate response variable of multiple predictors one may simply write

\begin{equation}\label{Eq22}
y = \alpha\*X + \varepsilon = \alpha_0+\alpha_1x_1+\alpha_2x_2+\cdots +\alpha_mx_m +\varepsilon, \quad \varepsilon \sim N(\mu, \sigma^2).
\end{equation}
It is clear that the response variable $y$ is normally distributed with mean $\mu$, and variance $\sigma^2$ and the linearity of the model
is apparent from the equation. One of the issues with this model is that, the assumptions about the data generating process are limited.
One remedy for this is to consider other types of distribution, and include a link function $g(.)$ relating the mean $\mu$, i.e.,

\begin{equation}
E(y) = \mu, \hspace{.3cm}
g(\mu) = \alpha\*X.
\end{equation}
In fact the typical linear regression model is a generalized linear model with a Gaussian distribution
and identity link function. To further illustrate the generalization, we may consider a distribution
other than the Gaussian for example
Poisson or a negative binomial distribution for a count data where the link function is natural log function.

\begin{equation}
g(y) = \alpha\*X + \varepsilon, \quad \varepsilon \sim D(\mu, \theta),
\end{equation}

\noindent where $D$ is any exponential family distribution. We can still generalize more to add nonlinear terms in the above equation namely

\begin{equation*}
\begin{array}{l}
g(\mu_i) = X_i\alpha + f_1(x_{1i}) + f_2(x_{2i})+f_3(x_{3i}, x_{4i}) +\cdots,\\
\end{array}
\end{equation*}

\noindent where $\mu_i=\mathbb{E}(y_i)$ and $y_i \sim$ \textit{an exponential or non-exponential family distribution}
and $f_js$ are any univariate or multivariate functions of independent variables called
smooth and nonparametric part of the model,
which mean that the shape of predictor functions are fully characterized
by the data as opposed to parametric terms that are defined by a
set of parameters like the parameter vector $\alpha$ in the linear part in which both to be estimated.

\noindent Having said that, the ordinary least square problem generalizes into

\begin{equation}
\mathcal{L}(\alpha)=(\|y-\mathbf{X}\alpha\|^{2}+\sum_{j=1}^{n}\lambda_j\*\int_{0}^{1}[{f^{(2)}_j}]^{2}(x))
\end{equation}

\noindent where $\lambda_j$, controlling the extent of penalization and establishes a trade-off
between the goodness of fit and the smoothness of the model.

\noindent On the other hand, these functions are represented using appropriate intermediate rank spline basis expansions of modest rank $k_j$, as

\begin{equation}
f_j(x) = \sum_{j=1}^{k_j}\beta_{ji}b_{ji}(x).\\
\end{equation}

\noindent Substituting the function under the integral sign with the above equation we get

\begin{equation}
\int_{0}^{1}[{f^{(2)}_j}]^{2}(x) = \beta^T\*S_j\*\beta
\end{equation}

\noindent where the right hand side is a quadratic form with respect to the known matrix $S_j$.

\noindent Collecting both the parametric and non-parametric coefficients
into a double parameter $(\alpha, \beta)$, we obtain

\begin{equation}
\mathcal{L}(\alpha, \beta)=(\|y-\mathbf{X}\alpha\|^{2}+\*\frac{1}{2}\sum_{j=1}^{n}\lambda_j\*\beta^T\*S_j\*\beta).
\end{equation}

\noindent Writing $S_\lambda = \sum_{j=1}^{n}\lambda_j\*S_j$ we get the more compact form

\begin{equation}
\mathcal{L}(\alpha, \beta)=(l(\alpha)+\*\frac{1}{2}\lambda_j\*\beta^T\*S_\lambda\*\beta).
\end{equation}

\noindent Let $\hat{\beta}$ be the maximizer of $\mathcal{L}$ and  $\mathcal{H}$ the negative Hessian of $\mathcal{L}$ at $\hat{\beta}$.
From a Bayesian viewpoint $\hat{\beta}$ is a posterior mode for $\beta$. The Bayesian approach views the smooth functions as
intrinsic Gaussian random fields with prior given by $N(0, S_\lambda^-)$,
where $S_\lambda^-$ is a Moore–Penrose orpseudoinverse of $S_\lambda$.
Furthermore in the large sample limit

\begin{equation}
\beta|y \sim N(\hat{\beta}, (\mathcal{H}+S_\lambda)^{-1}).
\end{equation}

\noindent Writing the density in (2.8) as $\mathcal{D}_g$, and the joint density of $y$ and $\beta$ as $\mathcal{D}(y, \beta)$, the Laplace approximation to the marginal likelihood for the smoothing parameters $\lambda$ and $\alpha$ is
$\mathcal{D}(\lambda, \alpha) = \mathcal{D}(y,\beta)/\mathcal{D}_g(\beta, y)$.
Finally, nested Newton iterations are used to find the values of $\log\lambda$ and $\alpha$; maximizing $\mathcal{D}(\lambda, \alpha$ ) and the corresponding $\hat{\beta}$. Wood, S. N. 2016 \cite{SWP1}

\section{Application of GAMs for fatal cases of Canada}
\noindent Let $y_k$ denotes the deaths reported on day $k$. We will study the Canada deaths in general and the provinces of Quebec, Ontario, and Alberta in particular cases.
Due to overdispersion seen in the deaths, we assume that $y_k$
follows a negative binomial distribution with mean $\mu$ and variance

$\mu+\mu^2/\theta$. Note that $NB(\theta, p)$, where $p= \mu/(\mu+\theta)$.

Then let

\begin{equation}
g(\mu) = f(t_k) + f_w(d_k) +f_b(d_k) + f_m(d_k),
\end{equation}

where $g$ is a link function log in our discussion,
$f$ is a smooth function of time, $t_k$, measured in days, $f_w$ is a zero mean cyclic smooth function
of day of the week $d_k \in \{1, 2, \cdots, 7\}$, set up so that $f_w^{(n)}(0) = f_w^{(n)}(7)$,
$f_b$ is a zero mean cyclic smooth function
of day of the biweek $d_k \in \{1, 2, \cdots, 14\}$, set up so that $f_b^{(n)}(0) = f_b^{(n)}(14)$,
$f_m$ is a zero mean cyclic smooth function
of day of the month $d_k \in \{1, 2, \cdots, 30\}$, set up so that $f_m^{(n)}(0) = f_m^{(n)}(30)$,
and  $n \in \{0, 1, 2\}$ denotes the order of the derivative in $f_w$, $f_b$ and $f_m$, see S. Wood 2020 \cite{SW}.
Based on the discussion and notations in section.1, $f$, $f_w$, $f_b$ and $f_m$ are basis functions representing the underlying death rate and the strong weekly, biweekly and monthly cycles seen in the data respectively.

\noindent From (2.8) we can easily compute the confidence intervals for each f and make inferences
about when the peak in f occurs. This can be done by executing gam function of mgcv library in R code.
The fitted models to the reported deaths in Canada, Quebec, Ontario, and Alberta are shown in figures 5, 6, 7 and 8 respectively. In the Canada case, we model the deaths with respect to day, day of week and day of month.
As the results of figure 5 shows, all the variables are statistically significant with $0.856$ as R-squared.

\begin{figure}
	\begin{center}
		\includegraphics[scale=.7]{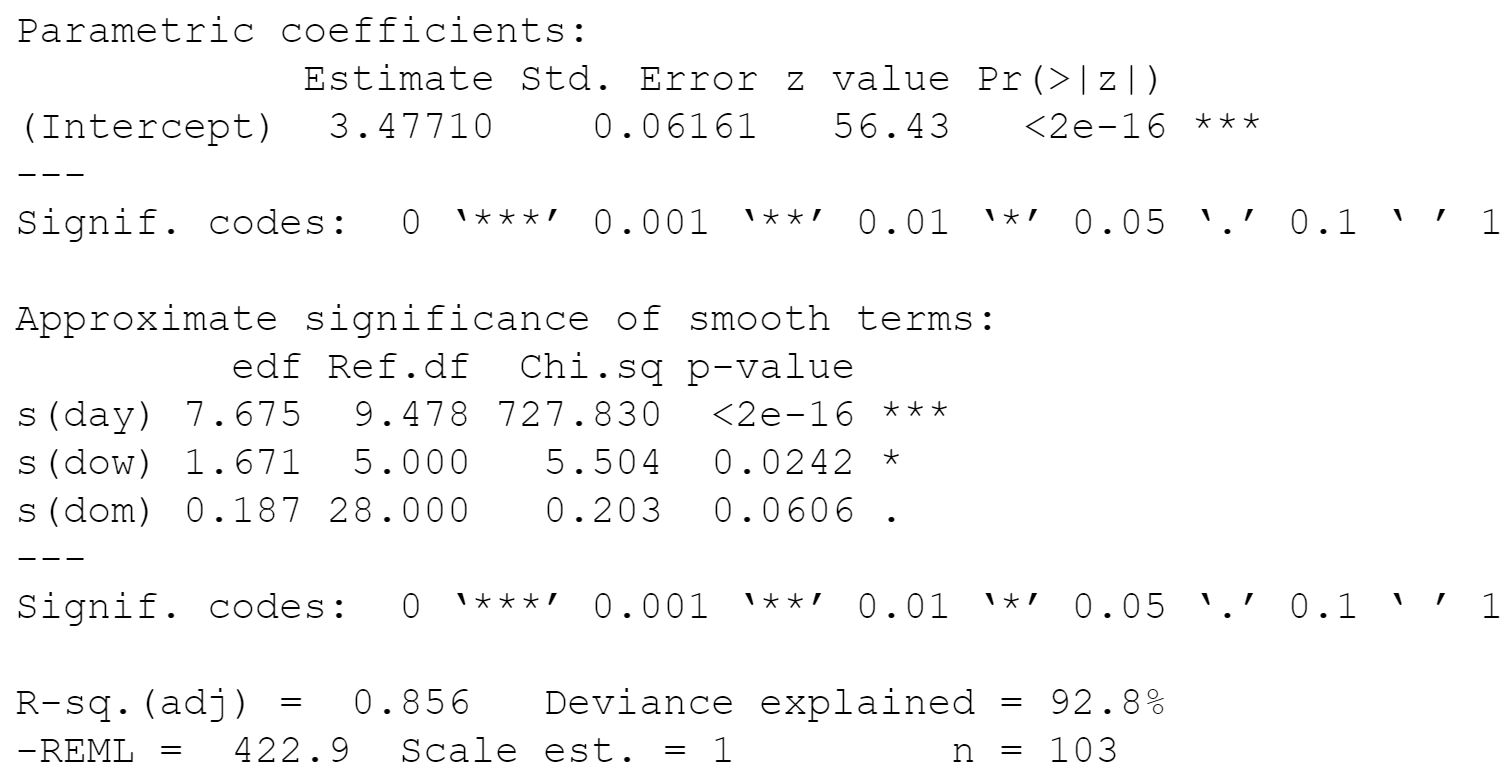}
		\caption{Results of the fitted model to the Canada deaths. Note that all variable are statistically
			significant.}\label{cansum.JPG}
	\end{center}
\end{figure}

\noindent For the deaths of Quebec, we see that the day, day of week
is an appropriate choice for the predicted variables.
As the results of figure 6 shows, all the variables are statistically significant with $0.75$ as R-squared.

\begin{figure}
	\begin{center}
		\includegraphics[scale=.7]{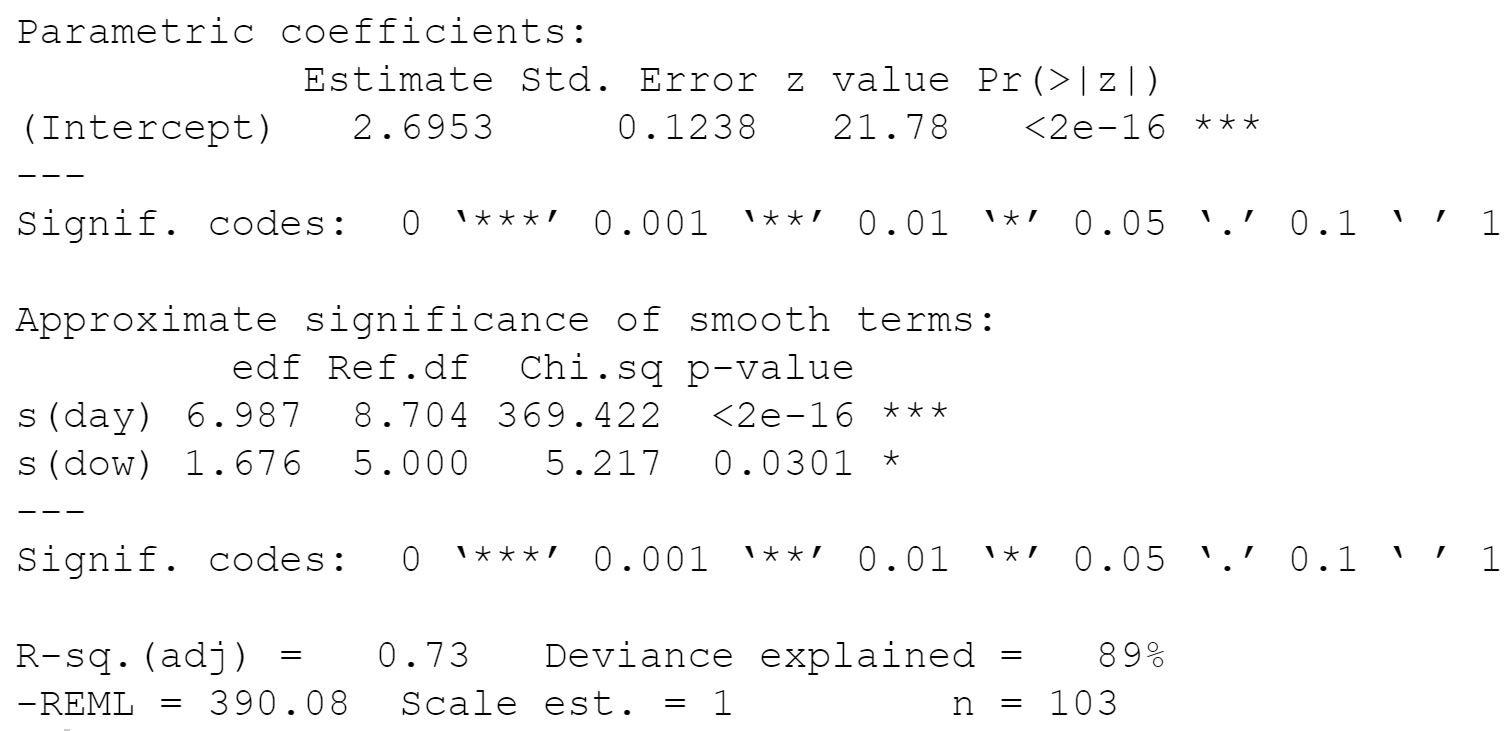}
		\caption{Results of the fitted model to the Quebec deaths. Note that all variable are statistically
			significant.}\label{quesum.JPG}
	\end{center}
\end{figure}

\noindent In the case of the Ontario we see that also the day, day of week and day of month
is an appropriate choice for the predicted variables.
As the results of Figure 7 shows, all the variables are highly significant with $0.82$ as R-squared.

\begin{figure}
	\begin{center}
		\includegraphics[scale=.7]{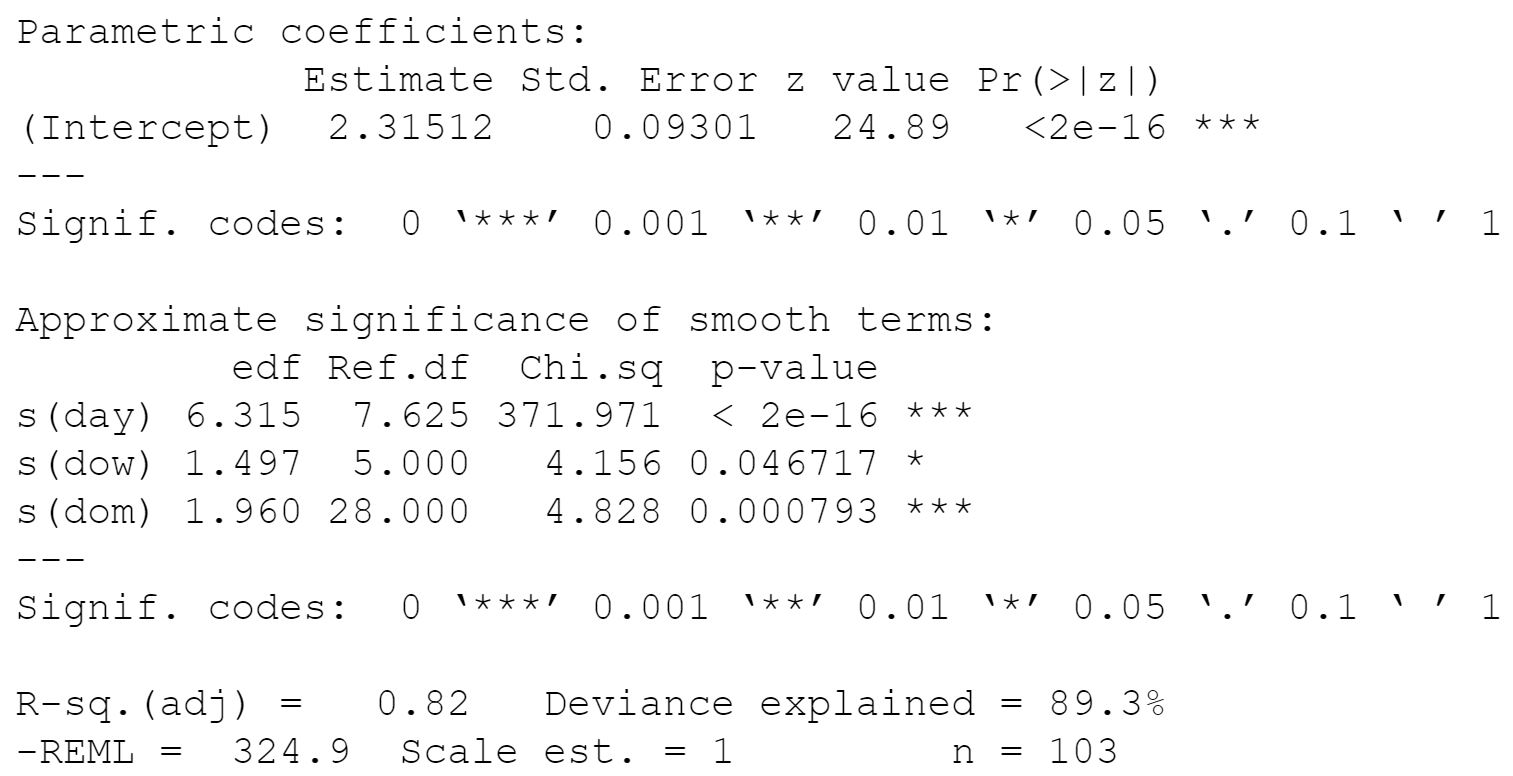}
		\caption{Results of the fitted model to the Ontario deaths. Note that all variable are highly
			significant.}\label{ontsum.JPG}
	\end{center}
\end{figure}

\noindent We figured out that the day, day of biweek as predictors for the Alberta case is an appropriate choice for the model.
As the results of figure 8 shows, all the variables are statistically significant with $0.722$ as R-squared.

\begin{figure}
	\begin{center}
		\includegraphics[scale=.7]{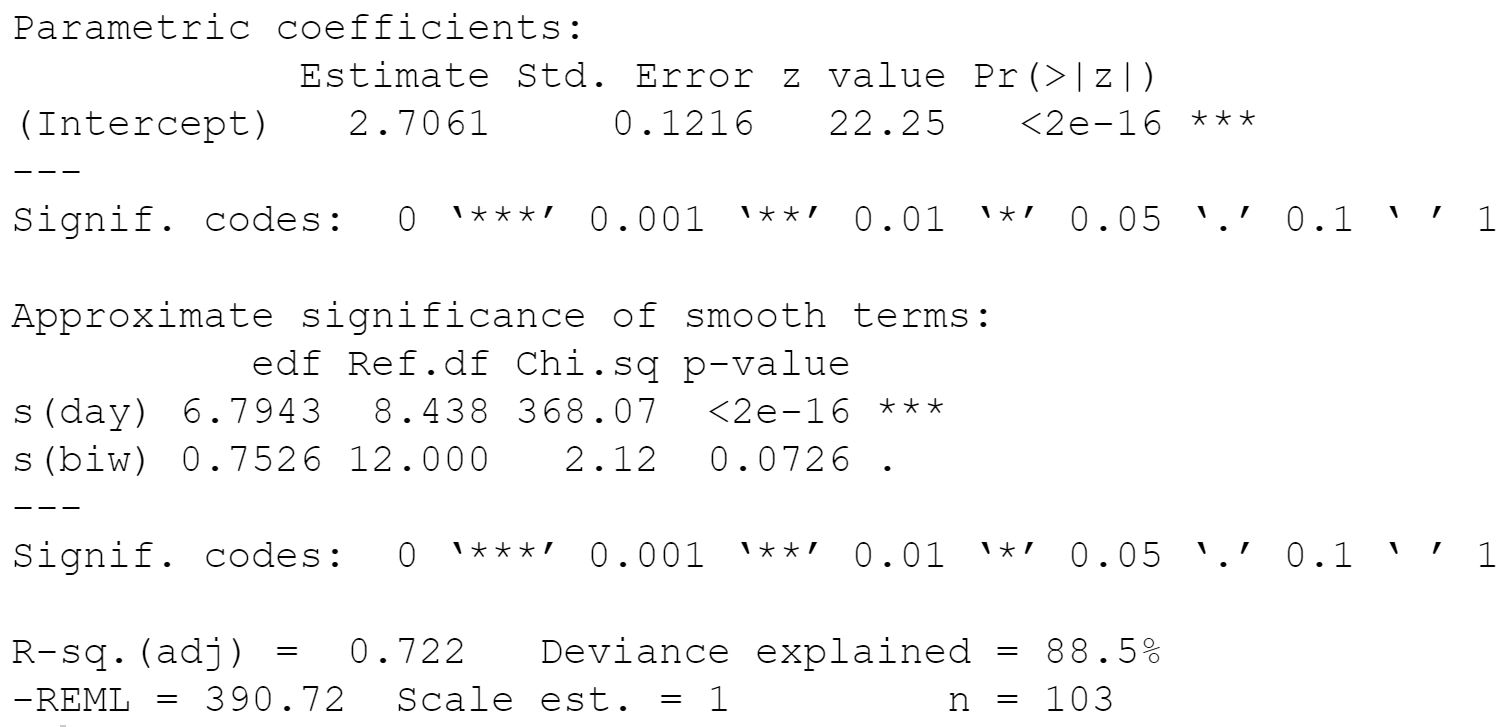}
		\caption{Results of the fitted model to the Alberta deaths. Note that all variable are statistically
			significant.}\label{albsum.JPG}
	\end{center}
\end{figure}

\noindent The next statistics is the results of the gam.check consists of Q-Q plot where for a good fit the data should lie on the red line. The second is the histogram of the residuals. In this plot, the histogram must be symmetric with respect to the line $x=0$.
Third plot is the Residuals vs. Linear predictors. This plot must be symmetric with respect to the line $y=0$.
Finally, the last plot is the Response vs. Fitted values. The more closer the data to the line $y=x$, the better the fitted model.
These plots for the corresponding data of Canada, Quebec, Ontario and Alberta are given in figures 9, 10, 11 and 12
respectively.

\begin{figure}
	\begin{center}
		\includegraphics[scale=1.1]{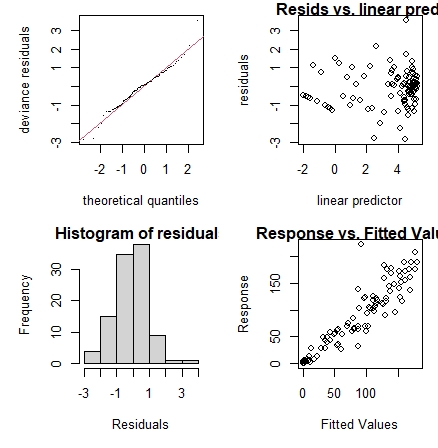}
		\caption{Canada: Q-Q plot, Histogram of residual, Residuals vs. Linear predictors and Response vs. Fitted values.}\label{cangam.jpeg}
	\end{center}
\end{figure}

\begin{figure}
	\begin{center}
		\includegraphics[scale=1]{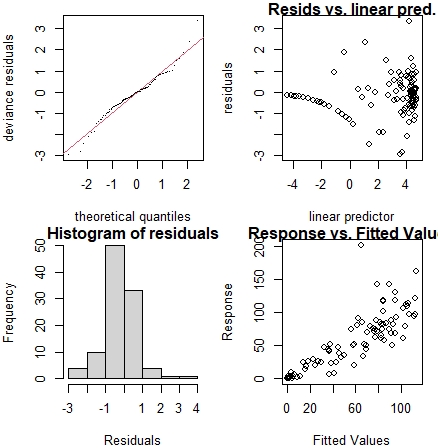}
		\caption{Quebec: Q-Q plot, Histogram of residual, Residuals vs. Linear predictors and Response vs. Fitted values.}\label{quegam.jpeg}
	\end{center}
\end{figure}

\begin{figure}
	\begin{center}
		\includegraphics[scale=1]{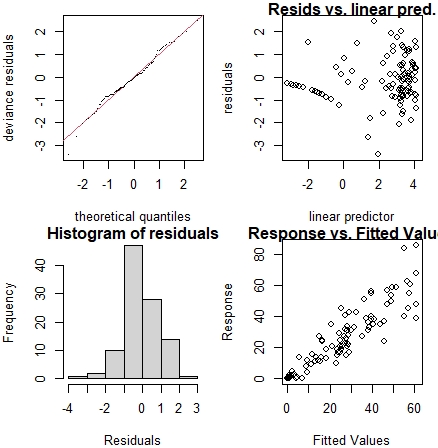}
		\caption{Ontario: Q-Q plot, Histogram of residual, Residuals vs. Linear predictors and Response vs. Fitted values.}\label{ontgam.jpeg}
	\end{center}
\end{figure}

\begin{figure}
	\begin{center}
		\includegraphics[scale=1]{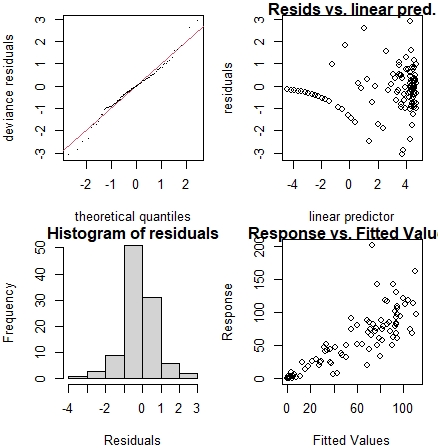}
		\caption{Alberta: Q-Q plot, Histogram of residual, Residuals vs. Linear predictors and Response vs. Fitted values.}\label{albgam.jpeg}
	\end{center}
\end{figure}

\noindent The other statistics of the model fits to the reported deaths in all cases are shown in figures 13, 14, 15 and 16 respectively.
The posterior modes (solid) and 95\% confidence intervals for the model functions as well as auto-correlation functions
and the deviance residuals against day for the Quebec and Alberta.

\begin{figure}
	\begin{center}
		\includegraphics[scale=1]{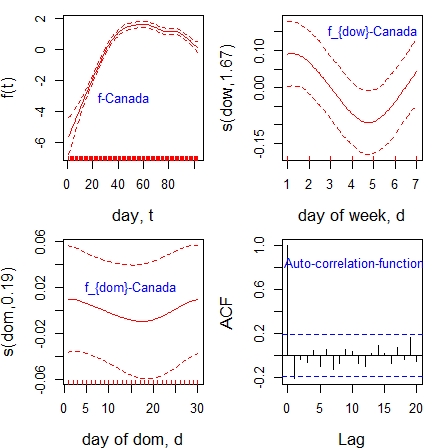}
		\caption{Posterior modes (solid) and 95\% confidence intervals for the model functions for the Canada
			as well as auto-correlation functions of the deviance residuals. Lag 1 is just significantly negative.}\label{cansm.jpeg}
	\end{center}
\end{figure}

\begin{figure}
	\begin{center}
		\includegraphics[scale=1]{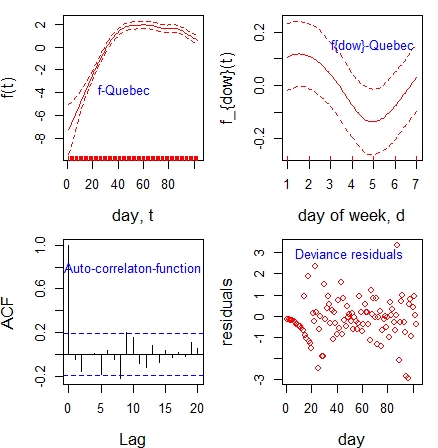}
		\caption{Posterior modes (solid) and 95\% confidence intervals for the model functions for the Quebec
			as well as auto-correlation functions of the deviance residuals and deviance residuals against day. Lag 7 is just significantly negative.}\label{quesm.jpeg}
	\end{center}
\end{figure}

\begin{figure}
	\begin{center}
		\includegraphics[scale=1]{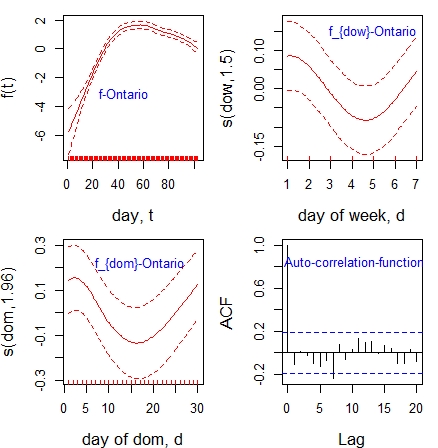}
		\caption{Posterior modes (solid) and 95\% confidence intervals for the model functions for the Ontario
			as well as auto-correlation functions of the deviance residuals. Lag 7 is just significantly negative.}\label{ontsm.jpeg}
	\end{center}
\end{figure}

\begin{figure}
	\begin{center}
		\includegraphics[scale=1.1]{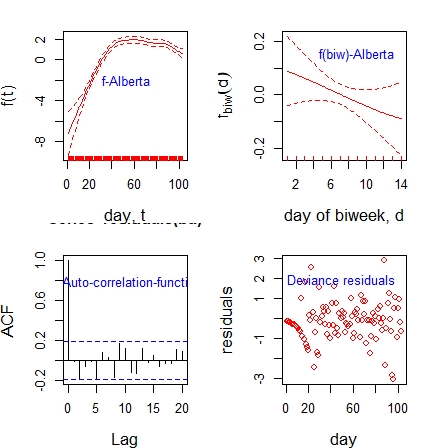}
		\caption{Posterior modes (solid) and 95\% confidence intervals for the model functions for the Alberta
			as well as auto-correlation functions of the deviance residuals against day. No significant lag.}\label{albsm.jpeg}
	\end{center}
\end{figure}
\section{Inference about the peak of the day of each four cases}

\noindent With gam model it is also straightforward to make inferences
about when the peak in f occurs. To this end, it is enough to use the model matrix by removing cyclic part, the coefficients and variance-covariance matrix of the model to estimate the model functions and $95\%$ confidence intervals.
To find the day of occurrence of the peak for each corresponding underlying death rate function, $f$, simply we generates multivariate normal random deviates using 'rmvn' function from 'mgcv' package in which it takes 3 arguments as number of simulations, the coefficients and variance-covariance matrix of the model. The results for all 4 cases are shown on the figures 17 and 18 respectively.

\begin{figure}
	\begin{center}
		\includegraphics[scale=.55]{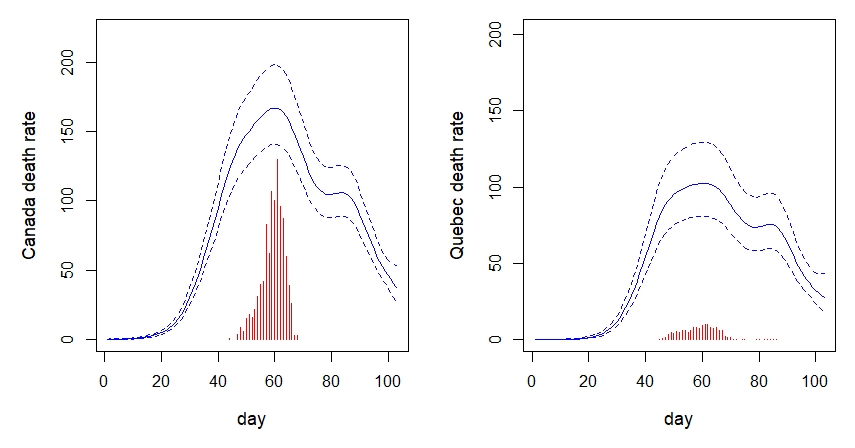}
		\caption{Underlying death rates for the Canada(left) and Quebec (right) with $95\%$ confidence intervals. Day 0 is March 17th 2020
			the Canada lock down. The scaled red bar chart illustrates the distribution
			of the location of the peak of the death rate curve, obtained by simulation from the approximate posterior
			for each model.}\label{canmedian1.jpeg}
	\end{center}
\end{figure}

\begin{figure}
	\begin{center}
		\includegraphics[scale=.45]{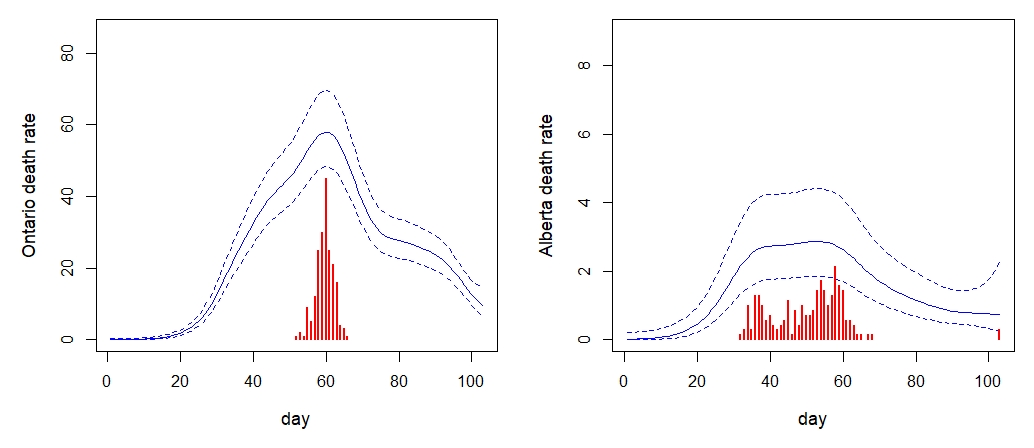}
		\caption{Underlying death rates for the Ontario(left) and Alberta (right) with $95\%$ confidence intervals. Day 0 is March 17th 2020
			the Canada lock down. The scaled red bar chart illustrates the distribution
			of the location of the peak of the death rate curve, obtained by simulation from the approximate posterior
			for each model.}\label{albmedian1.jpeg}
	\end{center}
\end{figure}

\section{Inference about the past fatal infection cases}
To obtain the model priors with auto-generated code and associated data is to simulate jagam from rjags library in GAMs.
We also load glm to improve samplers for GLMs.
This is useful for inference about models with complex random effects structure. The new mgcv function jagam is designed to be called in the same way that the modeling function gam would be called. That is, a model formula and family object specify the required
model structure, while the required data are supplied in a data frame or list or on the search
path. However, unlike gam, jagam does no model fitting. Rather it writes JAGS code to
specify the model as a Bayesian graphical model for simulation with JAGS, and produces a
list containing the data objects referred to in the JAGS code, suitable for passing to JAGS
via the rjags function jags.model (Plummer 2016)\cite{PL}.
To infer the sequence of past fatal infections one needs to produce the
observed sequence of deaths. Verity et al. 2020 \cite{VE} show that the distribution of time from onset of symptoms to death for
fatal cases can be modelled by a gamma density with mean 17.8 and variance 71.2 (s.d. 8.44).
Let $f_v(t)$ be the function describing the variation in the number of eventually fatal cases over time.
Let $\mathbf{B}$ be the lower triangular square matrix of order $n$, describing the onset-to-death gamma density, given above where $n$ is the number of day of pandemic under consideration. Then we have $\mathbf{B}_{ij}=\gamma(i-j+1)$  if
$i \geq j$ and 0 otherwise. Now $\mathbf{B}f_v = h$, where $h(k)$ is the expected number of deaths on day $k$. Then
$\log(f_v(k))$ can be represented using an intermediate rank spline, again with a cubic spline penalty. We can then employ exactly the
model of the previous section but setting $f(k) = \log(h(k))$. The only difference is that we need to infer $f_v$ over a
considerable period before the first death occurs where 15 days is clearly sufficient given the form of deaths data.
After executing the jagam code for the data frame of deaths, day, day of week, day of biweek and day of month variables - note that we
have used different combination of the above variables to see which one is appropriate of Canada deaths in general and Quebec, Ontario and Alberta in particular cases. Jagam code with Poisson family distribution produces a model containing all the priors. Then by adding
the matrix $\mathbf{B}$ and a bit of extra regularization to the output we pass the model to jags.model function
JAGS is Just Another Gibbs Sampler. It is a program for the analysis of Bayesian models
using Markov Chain Monte Carlo (MCMC) simulation.
Then by passing the result to the jags.samples function with the parameters of thin=300 and iteration=1000000 we get the values of $\theta$, $\rho$, and the Monte carlo array $b$ of 3 dimension with the first dimension equals to the number of parameters in the gam model. As in the gam model, we take the data in the first component of the jags model as the model.matrix $\mathbf{X}$ and the data in first 2 dimension of the $b$ to simulate the fatal infection profiles $f = \exp(Xb)$ and get all the necessary statistics such as median with confidence intervals, the peak points of the median profile, the squared second difference of the median infection profile on the log scale which is proportional to the smoothing penalty and the absolute gradient of the
infection profile. See Wood, S. N. 2016 \cite{SWP2} These are depicted in figure 19.

\begin{figure}
	\begin{center}
		\includegraphics[scale=0.7]{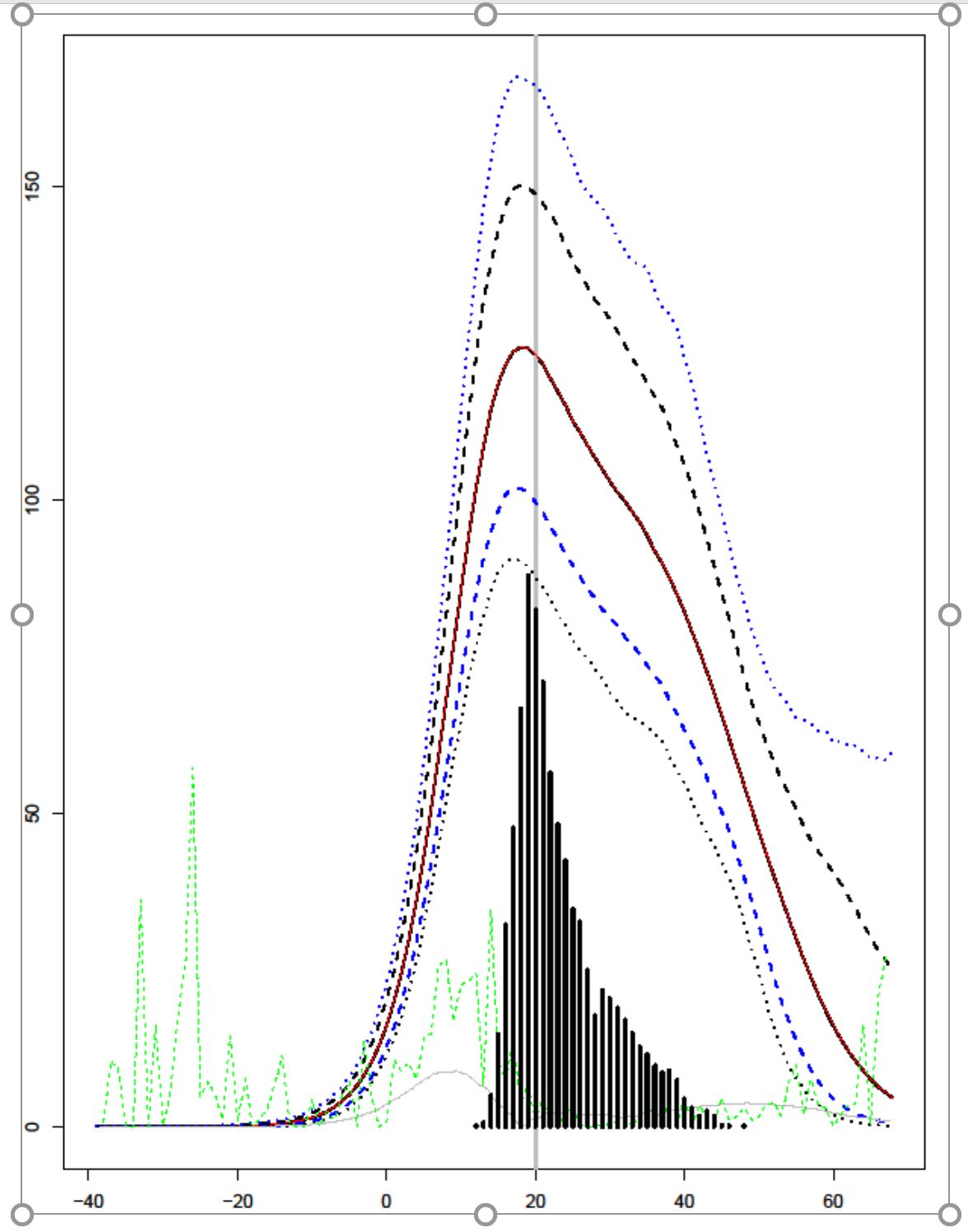}
		\caption{The inferred Canada fatal infection profile over time (day 0 is March 17th 2020). The red
			curve is the posterior median, the dashed curves delimit an $80\%$ confidence interval and the dotted curves
			a $95\%$ confidence interval. The scaled black bar chart shows the posterior distribution of the day of peak fatal
			infection. The dashed grey curve is proportional to the squared second difference of the median infection profile on the log scale
			which
			is proportional to the smoothing penalty. The green curve is proportional to the absolute gradient of the
			infection profile.}\label{canmedian2.JPG}
	\end{center}
\end{figure}

\clearpage

\textbf{Acknowledgement. }
I am very grateful to Simon S. Wood as his paper \cite{SW} was very inspirational to my work.

\bibliographystyle{amsplain}

\end{document}